\documentstyle[prl,aps,floats,psfig,preprint]{revtex}

\draft

\begin{document}

\title{
Charge correlations and dynamical instabilities 
in the multifragment emission process
}
\author
{
 L.G. Moretto,
 Th. Rubehn,
 L. Phair,
 N. Colonna,\cite{1}
 and G.J. Wozniak
}
\address{
Nuclear Science Division,
Lawrence Berkeley National Laboratory,
}
\address{
University of California, Berkeley, California 94720
}
\author{
D.R. Bowman,\cite{2}  
G.F. Peaslee,\cite{3}
N. Carlin,\cite{4} 
R.T. de Souza,\cite{5}
C.K. Gelbke,  
W.G. Gong,\cite{6}
Y.D. Kim,\cite{7}
M.A. Lisa,\cite{8} 
W.G. Lynch,  
and 
C. Williams
}
\address{National Superconducting Cyclotron Laboratory and Department of  
Physics and  
Astronomy, \\
Michigan State University, East Lansing, MI 48824}
%
%

\maketitle

\begin{abstract}
A new, sensitive method allows one to search for the 
enhancement of events with nearly equal-sized fragments
as predicted by theoretical calculations based
on volume or surface instabilities.
Simulations have been performed to investigate the 
sensitivity of the procedure. 
Experimentally, charge correlations of intermediate mass 
fragments emitted from heavy ion reactions at intermediate 
energies have been studied. 
No evidence for a preferred breakup into equal-sized fragments
has been found.
\end{abstract}

\pacs{PACS number(s): 
      25.70.Pq}	    

\narrowtext

In recent years, multifragmentation of nuclear systems
has been extensively studied, and many efforts have been made
to clarify the underlying physics \cite{Mor93}.
It has been suggested that fragment production can be
related to the occurrence of instabilities in the 
intermediate system produced by heavy ion collisions
\cite{Bro89,Mor92,Bau92,Gro92,Sou93,Xu93,Pha93a,Gla93,Bor93,Pal94,Cho95,Han95}.
In particular, two kinds of instabilities are extensively
discussed in the literature: volume instabilities of a 
spinodal type (see e.g. Ref.~\cite{Cho95}) 
and surface instabilities \cite{Mor92}. 
Spinodal instabilities are associated with the transit 
of a homogeneous fluid across a domain of negative pressure,
where the homogeneous fluid becomes unstable and breaks up 
into droplets of denser liquid.
Surface instabilities can be subdivided into Rayleigh or cylinder
instabilities which are responsible for the decay
of shapes like long necks or toroids \cite{Bro89},
and sheet instabilities which cause the decay of bubbles or 
disklike structures \cite{Mor92}. 
A variety of models have predicted the formation
of these exotic geometries which may develop after the initial 
compression of nuclei in the early stage of the collision
for both symmetric and asymmetric systems
\cite{Mor92,Bau92,Sou93,Xu93,Bor93,Pal94,Cho95,Han95}.
Although the scenarios and the models vary, breakup 
into several {\it nearly equal-sized} fragments
has been predicted for both kinds of
instabilities.
Thus it would be interesting to search model-independently
for this signal.
In this paper, we examine the signatures of a breakup 
configuration which would decay into a number of nearly 
equal-sized fragments by investigating charge correlations
from both experimental data and simulations. 

We have experimentally studied the reactions Xe+Cu at 
50 MeV/nucleon. The measurements were performed
at the National Superconducting Cyclotron Laboratory
of Michigan State University using the Miniball
\cite{Sou92} and a Si-Si(Li)-plastic forward array
\cite{Keh92}.
Detailed information on the experiment can be 
found in Ref.~\cite{Bow92}.

For comparison, and in order to determine the sensitivity 
of our analysis, Monte Carlo calculations have been performed. 
The created events obey two conditions: the sum charge of
all fragments is conserved within an adjustable accuracy, 
and a fragment is produced according to the probability 
resulting from the experimental finding, that 
the charge distributions for intermediate mass fragments 
(IMF) are nearly exponential functions \cite{Pha95}:
\begin{equation}
P_n(Z) \propto \exp(-\alpha_n Z).
\label{eq0}
\end{equation}
Experimentally, a variation of the parameter $\alpha_n$
between 0.2 and 0.4 has been reported for extreme 
cuts on the transverse energy $E_t$. 
In our simulations, we have chosen 
a fixed value of 0.3. The size of the decaying source
was chosen to be equal to the sum charge of xenon
and copper, $Z_{source} = 83$.
Events with equal sized fragments of charge $Z_{art}$ 
were randomly added with probability $P$
to simulate a dynamical breakup of the system
into nearly equal-sized pieces. 
Furthermore, the simulation allows one to smear out the 
charge distributions of the individual fragments of 
such an event according to a Gaussian distribution. 
This smearing of the charge distribution not
only accounts for the width of the distribution 
due to the formation process {\it per se}, 
but also for the probable sequential decay of the 
primary fragments (i.e. the evaporation of light charged 
particles). 
In the following, the full width at half-maximum of 
this distribution is denoted by $\omega$.
We have demanded that at least 75\% of the 
total available charge is emitted according to
Eq.~\ref{eq0}; i.e. the production of particles was
stopped in the simulation once this percentage had been
reached. 
We note that in this simple approach the 
transverse energy $E_t$ has not been simulated.

First, we investigate two particle correlations.
Both the experimental and the simulated events 
have been analyzed according to the following 
method.
The two particle charge correlations  
are defined by the expression
\begin{equation}
 \left.\frac{Y(Z_1,Z_2)}{Y'(Z_1,Z_2)}\right|_{E_t,N_{IMF}} = 
 \left.C[1 + R(Z_1,Z_2)]\right|_{E_t,N_{IMF}}.
\label{eq1}
\end{equation}
Here, $Y(Z_1,Z_2)$ is the coincidence yield of two particles
of atomic number $Z_1$ and $Z_2$ in
an event with $N_{IMF}$ intermediate mass fragments
and a transverse energy $E_t$ 
(for the definition of the latter, see Ref.~\cite{Pha93}).
The background yield $Y'(Z_1,Z_2)$ is constructed by mixing
particle yields from different coincidence events selected by
the same cuts on $N_{IMF}$ and $E_t$. 
The normalization constant $C$ is chosen 
to equalize the  integrated yields of $Y$ and $Y'$. 

\begin{figure}[htb]
 \centerline{\psfig{file=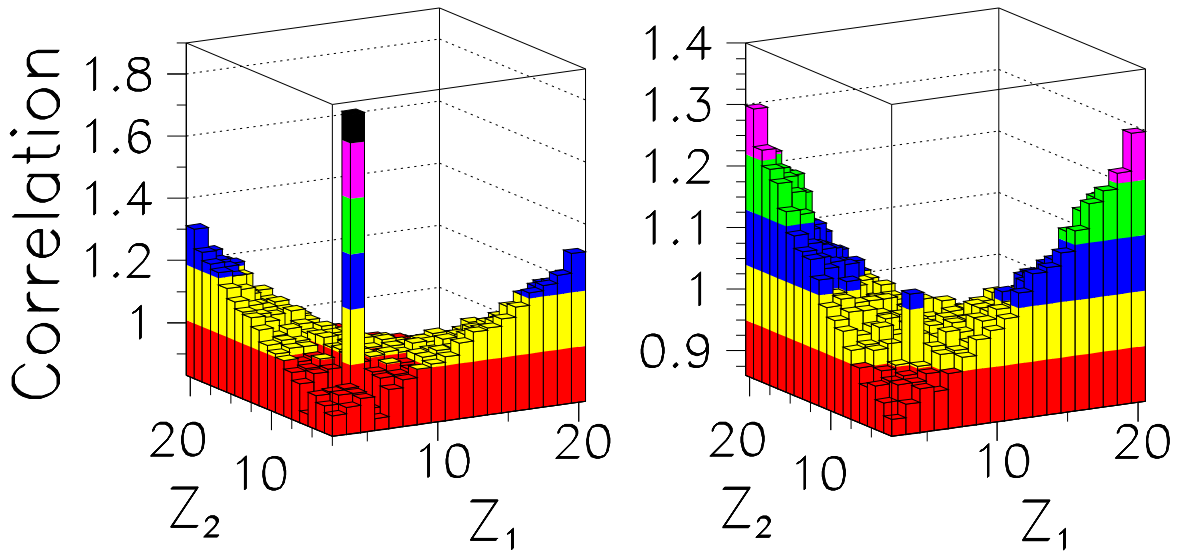,height=8.5cm}}
 \caption[]{
 Two particle charge correlations of intermediate mass fragments from 
 simulations investigating events with $N_{IMF} = 6$ and a 
 source size of 83.
 Randomly, 1\% (left panel) and 0.1\% (right panel) of the 
 events were chosen to have equal sized fragments ($Z_{art}$ = 6). 
 }
 \label{fig1}
\end{figure}

To demonstrate the sensitivity of our method 
to breakup configurations producing equal-sized fragments,
we show in Fig.~\ref{fig1} the results of simulations
for the case $N_{IMF} = 6$. Here, 1\% of the events 
consist of fragments which all have the size $Z_{art} = 6$.
The peak produced by these fragments is clearly visible,
even if we decrease the yield of equal-sized fragments to
only 0.1\%.

Historically, charge correlations have often been investigated
by using Dalitz plots \cite{Kre93,Gua96}.
However, this technique does not provide a sensitive 
tool to search for an {\it enhanced} breakup into several
nearly equal-sized fragments, since the ``background'' is ignored. 
The Dalitz analysis is equivalent to studying {\it only} 
the numerator of the charge correlation function and will 
reflect only charge conservation in our search for relatively 
small enhancements.
Thus, the strong signal shown in Fig.~\ref{fig1} does not 
show up in a Dalitz plot.
Furthermore, the Dalitz plots and the charge distributions 
of the three largest fragments from Ref.~\cite{Gua96}, which 
claim to be a signature for an enhanced breakup into nearly 
equal-sized fragments, can be reproduced trivially by our 
simple simulation. Thus, these ``signals'' reflect only
the background produced by charge conservation.

\begin{figure}[htb]
 \centerline{\psfig{file=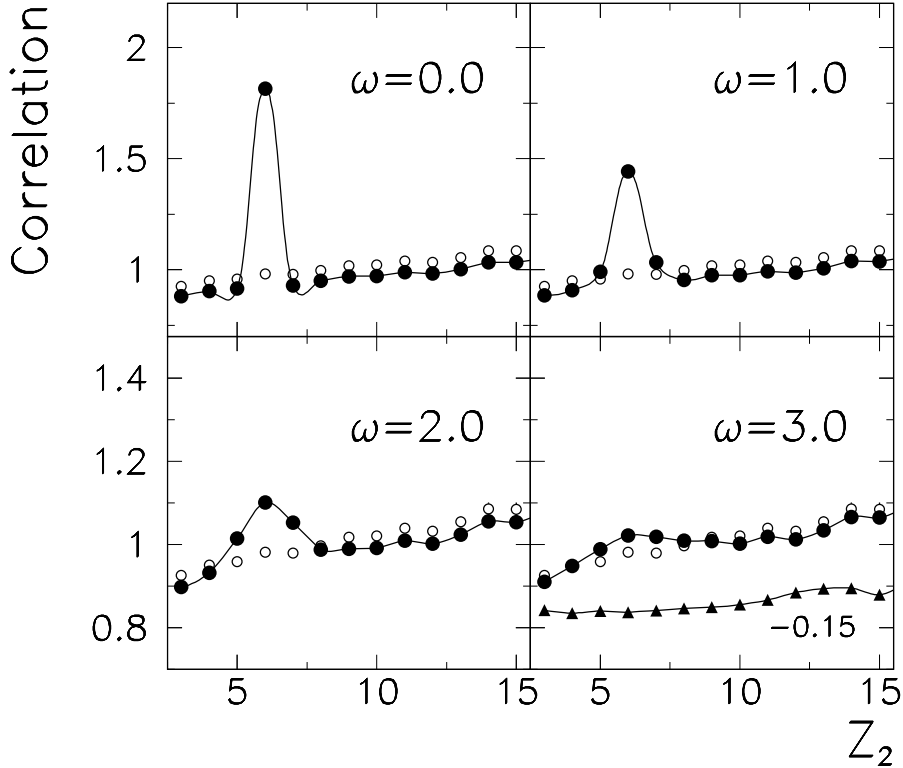,height=11cm}}
 \caption[]{
 Two particle charge correlations resulting from 
 simulations for $Z_1 = 6$ as a  function of the fragment charge 
 $Z_2$ for different values of the width of the charge 
 distribution $\omega$.
 Randomly, 1\% of the events were chosen to have nearly equal sized
 fragments (full circles). 
 For comparison, we have also plotted 
 a calculation where no additional events with 
 equal-sized fragments have been added (open circles).
 Experimental results for the case $N_{IMF}=6$ are shown in 
 the right lower panel (full triangles). For clarity, 
 these values are vertically shifted by a value of -0.15. 
 The error bars are smaller than the size of the symbols.
 }
 \label{fig2}
\end{figure}

\begin{figure}[htb]
 \centerline{\psfig{file=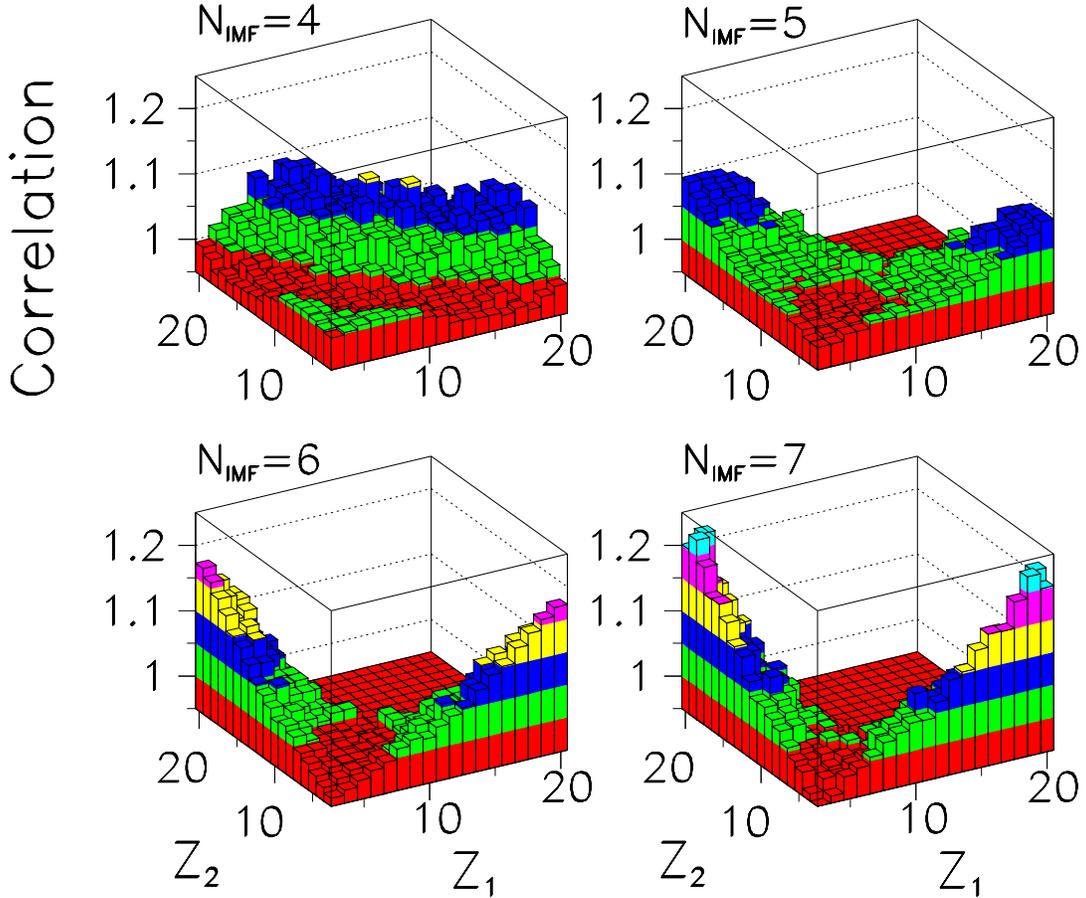,height=13cm}}
 \caption[]{
 Experimental two particle charge correlations for the 
 reaction Xe+Cu at
 50 MeV/nucleon. The different figures correspond to 
 $N_{IMF}$ cuts between 4 and 7.}
 \label{fig3}
\end{figure}

The magnitude of the peak shown in Fig.~\ref{fig1}
depends not only on
the yield, but also on the width of the charge distribution
of the nearly equal-sized fragments. In Fig.~\ref{fig2},
we show the correlation functions (solid circles)
for different widths $\omega$
and for $Z_1 = 6$. 
For comparison, we have plotted the results 
of a calculation (open circles) where no additional events with 
equal-sized fragments have been added: 
As expected, a dependence of the size of the peak on the 
smearing can be observed which limits the sensitivity 
of the two particle correlation functions to an enhancement 
of events where the charge distribution is relatively narrow. 
However, for the narrow widths
predicted as e.g. in Ref.~\cite{Xu93},
a clear signal should be visible in the experimental data.
The same analysis used for the simulation 
has been applied to the experimental data. 
In Fig.~\ref{fig3}, we show the
results for the reaction Xe+Cu at 50 MeV/nucleon
for different cuts in the intermediate 
mass fragment multiplicity $N_{IMF}$. 
A top 5\% cut on $E_t$ has been applied in order to
select central events.
With higher fragment multiplicity the distribution
peaked along the line $Z_1 + Z_2 \approx 30$
changes into a distribution peaked
at values where one fragment is heavy and 
its partner is light.
However, an enhanced signal for breakup into nearly equal-sized
fragments (a signal appearing along the diagonal)
was not observed in {\it any} of the $N_{IMF}$ bins.
As an example, we show in Fig.~\ref{fig2} the 
experimental two particle correlation function vs. $Z_2$
for $N_{IMF}=6$ and $Z_1=6$ (triangles).

Furthermore, we have investigated the correlation functions 
obtained by our simulations without enhanced breakup for several 
intermediate mass fragment multiplicities.
The evolution of the shape of the distribution 
with increasing values of $N_{IMF}$ is very similar
to that observed in the
experimental data of Fig.~\ref{fig3}.
Simulations with different system sizes show that 
the charge correlations decrease as $Z_{source}$
increases; this can be attributed to the 
definition of an IMF ($3 \le Z_{IMF} \le 20$)  
relative to $Z_{source}$.
We have also performed calculations 
using a percolation code and have observed a
dependence similar to that presented in Fig.~\ref{fig3}.
In order to study whether the evolution of the
distributions' shape with multiplicity is only
due to charge conservation, we have investigated
the breakup of an integer number $Z_0$ (chain) into
$n$ pieces. The latter were produced by 
$(n-1)$ random breaks of the bonds. 
The calculated two particle correlation
functions for different multiplicities $n$ 
have a similar evolution of the distribution
with $n$ as shown in Fig.~\ref{fig3}. 
These findings indicate that the observed 
experimental evolution of the shape of the two particle 
charge correlation distribution with fragment multiplicity
is due to the limited number of 
possibilities to create fragments if both the sum charge 
and the number of fragments are fixed.
A signal of enhanced emission will sit on top
of such a background.

To search for weak signals of events with
nearly equal-sized fragments, and in the hope of increasing
the sensitivity of the method, we have investigated 
higher order charge correlations
\cite{note1}.
This quantity is defined by the expression:
\begin{equation}
 \left.\frac{Y(\Delta Z,\langle Z \rangle)}{Y'(\Delta Z, \langle Z \rangle)}
 \right|_{E_t,N_{IMF}} = 
 \left.C[1 + R(\Delta Z, \langle Z \rangle)]\right|_{E_t,N_{IMF}}.
\label{eq2}
\end{equation}
Here, $\langle Z \rangle$ denotes the average fragment charge of the event,
$\langle Z \rangle = \sum_{i=1}^{N_{IMF}} Z_i/N_{IMF}$, and
$\Delta Z$ is the standard deviation, defined by 
$\Delta Z$ = 
$\sqrt{(N_{IMF}-1)^{-1} \sum_{i=1}^{N_{IMF}}(Z_i-\langle Z \rangle)^2}$.
The normalization constant $C$ and the yields are 
defined according to Eq.~\ref{eq1}. 
The denominator for the background yield 
$Y'(\Delta Z, \langle Z \rangle)$ is obtained by constructing 
``pseudo events'' where one fragment is selected from each of 
the previous $N_{IMF}$ events of the same event class 
(same $E_t$ range). 

\begin{figure}[htb]
 \centerline{\psfig{file=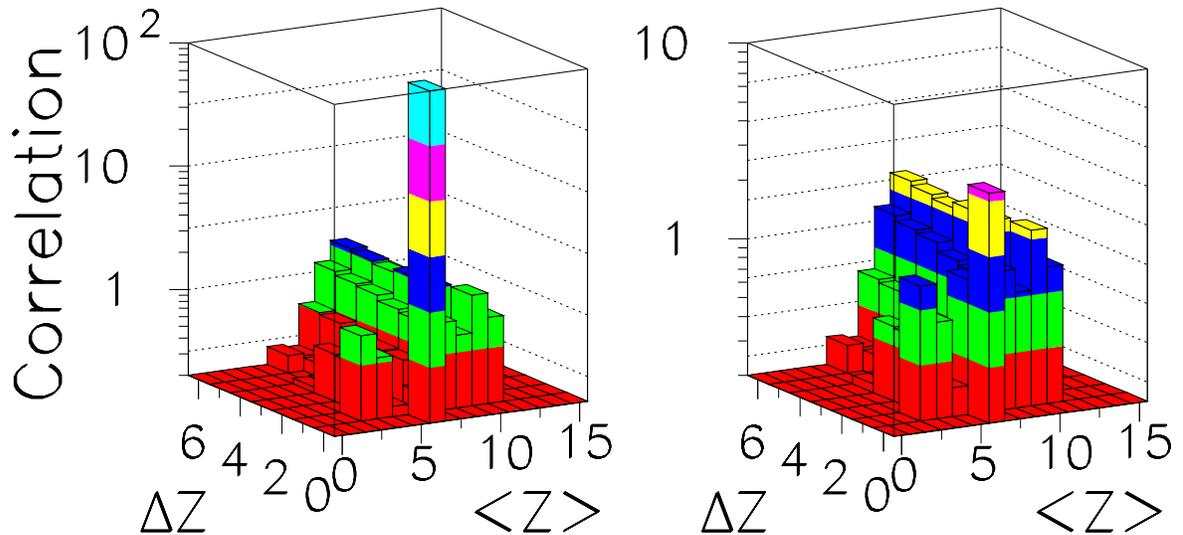,height=8.5cm}}
 \caption[]{
 Higher order charge correlations 
 from the simulations for $N_{IMF}=6$. 
 Randomly, 0.1\% of the events were chosen to have equal-sized
 fragments; A width of $\omega = 0$ has been chosen (left panel).
 On the right panel, the results for a width $\omega = 2$ are 
 presented. Note the logarithmic scale of the correlation axis.}
 \label{fig4}
\end{figure}

In Fig.~\ref{fig4}, we show the results of an analysis
investigating higher order charge correlations. The same
simulation which has already been shown in Fig.~\ref{fig1} 
was used.  
Here, only 0.1\% of the events were chosen to have fragments with 
equal size. We show two cases with a width of $\omega = 0$ and
$\omega = 2$, respectively. The comparison between the two particle 
and the higher order charge correlation functions for the same simulations
using $\omega = 0$ shows an enhancement of $\sim$20\% for the first
case (right panel of Fig.~\ref{fig1}) 
while the signal in the second case exceeds the
``background''   by roughly a factor of 100
(left panel of Fig.~\ref{fig4}).
Since the yield in the $\Delta Z = 0$ bin increases 
dramatically with {\it any} enhancement of events with equal-sized
fragments, it should be sufficient to examine
this bin only;
the correlations at higher values of $\Delta Z$ 
represent the ``background''. 

\begin{figure}[htb]
 \centerline{\psfig{file=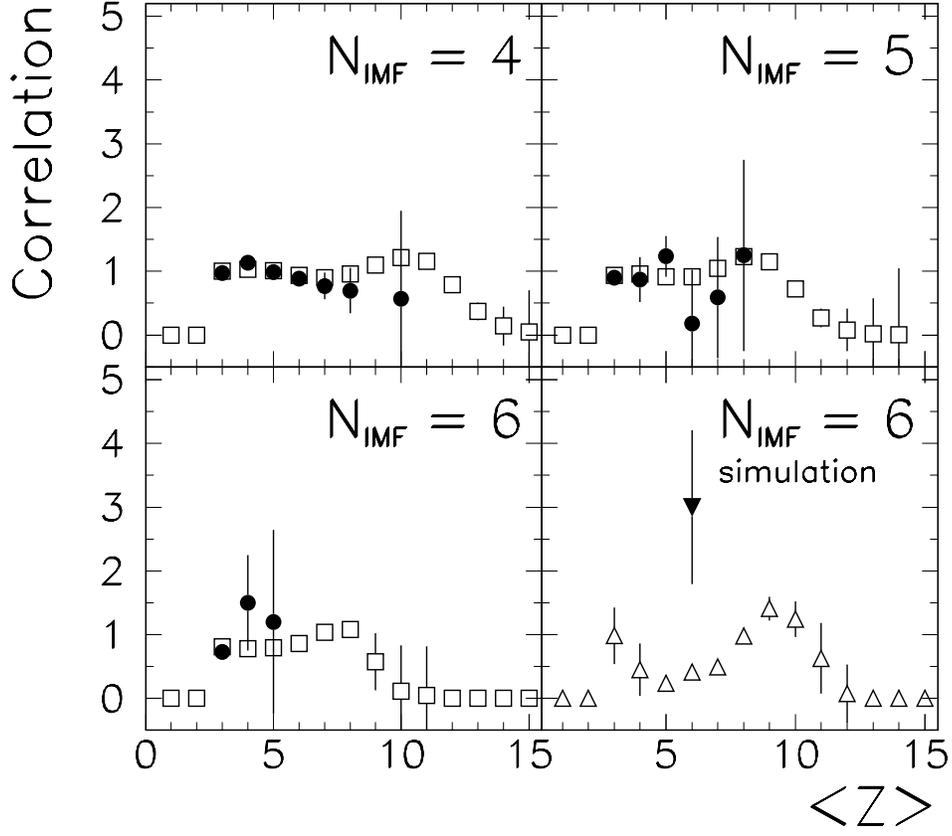,height=12cm}}
 \caption[]{
 Higher order charge correlations  
 for the reaction  Xe+Cu at 50 MeV/nucleon for $4 \le N_{IMF} \le 6$. 
 For comparison, we show the results of the simulation
 for  $N_{IMF}=6$ and $\omega = 2$ (lower right panel).
 The full symbols indicate the 
 events where $\Delta Z = 0$, the open symbols show the 
 ``background'' defined by $\Delta Z > 0$.
 For further details, see text.
 }
 \label{fig5}
\end{figure}

We have analysed our experimental data and determined the higher 
order charge correlation functions. The results are shown in 
Fig.~\ref{fig5} for the reaction Xe+Cu at 50 MeV/nucleon.
For comparison, we also show the results of the simulation
already shown in the right panel of Fig.~\ref{fig4} for 
$N_{IMF}=6$, $P = 0.1\%$, and $\omega = 2$.
As discussed above, we compare the correlation values for 
$\Delta Z = 0$ (solid symbols) with the data obtained 
for $\Delta Z > 0$ (open symbols).
No signals are observed that can be attributed
to an enhanced production of 
nearly equal-sized fragments.    
This results in an upper limit of breakup events with nearly equal-sized 
fragments of less than 0.05\% if we assume a width $\omega < 3$.
Furthermore, we have experimentally studied the reactions
Ar+Au at 50 and 110 MeV and Xe+Au at 50 MeV/nucleon 
\cite{Bow92,Pha93}:
Similar results have been obtained for these systems
albeit with significantly poorer statistics.
Thus, the Xe+Cu system has been used to establish
an upper limit and to explore the utility of this 
novel technique.

In conclusion, we have investigated two particle and higher 
order charge correlation functions of multifragment decays to search
for the enhanced production of nearly equal-sized 
fragments predicted in several theoretical works. 
Two particle charge correlation functions are sensitive
even to those enhancements from events where a number of 
equal-sized fragments might be accompanied by heavy partners,
as could be expected from the breakup of a necklike structure. 
While the higher order charge correlations are 
not sensitive to necklike emission, they do provide
an even more sensitive tool for identifying enhancements
associated with the emission of all nearly equal-sized fragments.
The analysis of experimental data for the reactions 
Xe+Cu and Xe+Au at 50 MeV/nucleon and Ar+Au at 50 and 
110 MeV/nucleon, however, shows no evidence for a 
preferred breakup into nearly equal-sized fragments.
Recently, two groups have reported experimental 
signatures of possible formations of non-compact 
geometries in the reactions $^{86}$Kr on $^{93}$Nb at 
$E/A$=65 MeV and Pb+Au at $E/A$=29 MeV,
respectively \cite{Sto96,Dur96}.
It would be very interesting to analyze 
these data using the new method  presented in this letter.

\bigskip    
This work was supported by the Director, Office of Energy Research,
Office of High Energy and Nuclear Physics, Nuclear Physics Division
of the US Department of Energy, under contract DE-AC03-76SF00098
and by the  National Science Foundation under  
Grant Nos. PHY-8913815, PHY-90117077, and PHY-9214992.


\begin{thebibliography}{References}{}

\bibitem[*]{1} Present address:
INFN-Sez. di Bari, 70126 Bari, Italy.

\bibitem[\dag]{2} Present address:
Chalk River Laboratories, Chalk River, Ontario K0J 1J0, Canada.

\bibitem[\ddag]{3} Present address:
Physics Department, Hope College, Holland, MI 49423.

\bibitem[\S]{4} Present address:
Instituto de Fisica, Universidade de Sao Paulo,  
C.P. 66318, CEP 04389-970, Sao Paulo, Brazil.

\bibitem[\parallel]{5} Present address:
Department of Chemistry, Indiana University,
Bloomington, IN 47405.

\bibitem[\P]{6} Present address:
Max-Planck-Institut f\"ur Physik, F\"oh\-ringer Ring 6, 
80805 M\"unchen, Germany.

\bibitem[**]{7} Present address:
Physics Department, Seoul National University,
Seoul, 151-742, Korea.

\bibitem[\dag\dag]{8} Present address:
Lawrence Berkeley National Laboratory, Berkeley, CA 94720.



\bibitem{Mor93}
L.G. Moretto and G.J. Wozniak,
Ann. Rev. Nucl. Part. Sci. {\bf 43}, 379 (1993),
and references therein.

\bibitem{Bro89}
U. Brosa, S. Grossmann, A. M\"uller, and E. Becker,
Nucl. Phys.{\bf\,A 502}, 423c (1989);
U. Brosa, S. Grossmann, and A. M\"uller,
Phys. Rep.{\bf\,197}, 167 (1990).

\bibitem{Mor92}
L.G. Moretto, K. Tso, N. Colonna, and G.J. Wozniak, 
Nucl. Phys. {\bf\,A 545}, 237c (1992);
Phys. Rev. Lett.{\bf\,69}, 1884 (1992).

\bibitem{Bau92}
W. Bauer, G.F. Bertsch, and H. Schulz,
Phys. Rev. Lett.{\bf\,69}, 1888 (1992).

\bibitem{Gro92}
D.H.E. Gross, B.A. Li, and A.R. DeAngelis,
Ann. Physik {\bf 1}, 467 (1992).

\bibitem{Sou93}
S.R. Souza and C. Ng\^{o}, 
Phys. Rev. C{\bf\,48}, R2555 (1993).

\bibitem{Xu93}
H.M. Xu {\it et al.}, Phys. Rev. C{\bf\,48}, 933 (1993).

\bibitem{Pha93a}
L. Phair, W. Bauer, and C.K. Gelbke, 
Phys. Lett. {\bf\,B 314}, 271 (1993).

\bibitem{Gla93}
T. Glasmacher, C.K. Gelbke, and S. Pratt,
Phys. Lett. {\bf\,B 314}, 275 (1993).

\bibitem{Bor93}
B. Borderie, B. Remaud, M.F. Rivet, and F. Sebille,
Phys. Lett. {\bf\,B 302}, 15 (1993).
 
\bibitem{Pal94}
S. Pal, S.K. Samaddar, A. Das, and J.N. De,
Phys. Lett. {\bf\,B 337}, 14 (1994).

\bibitem{Cho95}
Ph. Chomaz, M. Colonna, A. Guanera, and B. Jacquot,
Nucl. Phys.{\bf\,A 583}, 305c (1995)
and references therein.

\bibitem{Han95}
D.O. Handzy {\it et al.},
Phys. Rev. C{\bf\,51}, 2237 (1995).

\bibitem{Sou92}
R.T. de Souza {\it et al.}, Nucl. Inst. Meth. {\bf A 311},
109 (1992).

\bibitem{Keh92}
W.C. Kehoe {\it et al.}, Nucl. Inst. Meth. {\bf A 311},
258 (1992).

\bibitem{Bow92}
D.R. Bowman {\it et al.}, Phys. Rev. C{\bf\,46}, 1834 (1992).

\bibitem{Pha95}
L. Phair {\it et al.}, Phys. Rev. Lett.{\bf\,75}, 213 (1995).

\bibitem{Pha93}
L. Phair {\it et al.}, Nucl. Phys.{\bf\,A 564}, 453 (1993).

\bibitem{Kre93}
P. Kreutz {\it et al.},
Nucl. Phys.{\bf\,A 556}, 672 (1993).

\bibitem{Gua96}
A. Guarnera, Ph. Chomaz, and M. Colonna,
in {\it Proc. of the 34th International
Winter Meeting on Nuclear Physics}, 
Bormio, Italy, 1996,
ed. by I. Iori
[Ricera Scientifica ed Educazione Permanente, 1996] 
(in press).

\bibitem{note1}
The term ``higher order charge correlations'' reflects
the fact that all fragments of one events are
taken into account.

\bibitem{Sto96}
N.T.B. Stone {\it et al.}, 
in {\it Proc. of the 12th Winter Workshop on 
Nuclear Dynamics}, Snowbird, Utah, 1996, 
ed. by W. Bauer and G.D. Westfall
[Plenum, 1996] (in press).

\bibitem{Dur96}
D. Durand {\it et al.}, submitted to Phys. Lett. B;
J.F. Lecolley {\it et al.}, submitted to Phys. Lett. B.


\end{thebibliography}
\end{document}